# Observation of self-amplifying Hawking radiation in an analog black hole laser

Jeff Steinhauer

*Department of Physics, Technion—Israel Institute of Technology, Technion City, Haifa 32000, Israel*

**It has been proposed that a black hole horizon should generate Hawking radiation. In order to test this theory, we have created a narrow, low density, very low temperature atomic Bose-Einstein condensate, containing an analog black hole horizon and an inner horizon, as in a charged black hole. We observe Hawking radiation emitted by the black hole. This is the output of the black hole laser. We also observe the exponential growth of a standing wave between the horizons. The latter results from interference between the negative energy partners of the Hawking radiation and the negative energy particles reflected from the inner horizon. We thus observe self-amplifying Hawking radiation.**

When the idea of a black hole was first proposed, it was thought that a black hole was featureless, and that anything entering the black hole was completely destroyed, disappearing into the massive singularity at the center of the black hole. Bekenstein then realized that a black hole has entropy and effective temperature[1]. Hawking then applied quantum field theory to the curved spacetime near a black hole, and found that the black hole should emit radiation with the temperature predicted by Bekenstein[2,3]. The energy emitted is accompanied by the creation of negative energy partner particles which fall into the black hole, reducing its energy. However, the Hawking radiation should be exceedingly weak for known black holes with masses of a few solar masses or greater, preventing the experimental confirmation of Hawking's prediction. On the other hand, small black holes created in the early universe might emit detectable Hawking radiation[4], or perhaps very small black holes could be created and detected in particle accelerators[5].

Unruh pointed out that a quantum fluid could be used to form an analogue black hole in the laboratory[6]. Specifically, a transition from subsonic flow to supersonic flow is analogous to a black hole event horizon, since sound traveling against the flow cannot exit the supersonic region. A Bose-Einstein condensate could serve as the quantum fluid[7-10]. A variety of other systems have been proposed, including superfluid $^3$He [11], an electromagnetic waveguide[12], ultracold fermions[13], a ring of trapped ions[14], slow light in an atomic vapor[15-17], light in a nonlinear liquid[18], and an exciton-polariton condensate[19]. When waves generated in water were allowed to scatter from an analog white hole horizon, mode conversion as required for the Hawking process was observed[20,21]. Also, nonlinear optical fibers containing horizons[22,23] were seen to emit photons[23], but this was likely due to effects other than Hawking radiation[24,25]. Our observation of Hawking radiation can be considered a quantum simulation, such as those in trapped ions[26-29], quantum photonics[30], optical lattices[31], and superconducting circuits[32].

Corley and Jacobson considered a system analogous to a charged black hole, in which the black hole event horizon is followed by an inner horizon[33]. If the dispersion relation has upward curvature, they asserted that the negative energy partner particles of the Hawking radiation will reflect from the inner horizon and return to the black hole horizon. The reflected particles will then stimulate additional Hawking radiation, forming a black hole laser. This idea was further developed quantitatively in works by the St Andrews and Orsay groups[34-37]. Ref. 36 studies the density correlations, a technique pioneered by the Trento-Bologna collaboration[38]. Other works have studied a black hole laser imbedded in a ring-shaped condensate[7,39,40].

We created the analog of a charged black hole in our previous work[41]. We found experimentally that a potential step (a "waterfall" potential) is an excellent technique for accelerating an atomic condensate to well above the speed of sound, as required for creating the horizons. In the years since that work, we have developed techniques for studying small populations of phonons, with the goal of observing Hawking radiation[42,43]. We have now reached that goal; our black hole laser shows self-amplifying Hawking radiation. Our previous work inspired additional theoretical studies[36,44,45] which in turn have helped us plan our current experiment and interpret our observations.

Fig. 1a shows the dispersion relation in the subsonic region to the right of the black hole horizon *BH*, as well as in the supersonic lasing region between *BH* and the inner horizon *IH*. Only the relevant branch is shown, corresponding to the quasiparticles which travel against the flow (to the right) in the fluid frame. The solid part of the curve indicates positive norm modes, whereas the dashed part indicates negative norm, negative energy modes. To the right of *BH*, there is only one positive norm mode with frequency $\omega$. This is the Hawking radiation *HR,* which is also indicated in the spacetime diagram, Fig. 1b [33]. In the lasing region however, there are two negative norm modes and one positive norm mode with the same $\omega$. Thus, the horizon allows for the mixing of the positive and negative energy modes. By considering the group velocity $d\omega/dk$ of the modes, we see that the $in_-$ and $in_+$ modes in the lasing region are moving toward the black hole horizon. It is seen that these modes exist due to the superluminal (upward curving) dispersion relation. The $P$ mode on the other hand, moves away from the black hole horizon. This is the negative norm "partner" of the Hawking radiation.

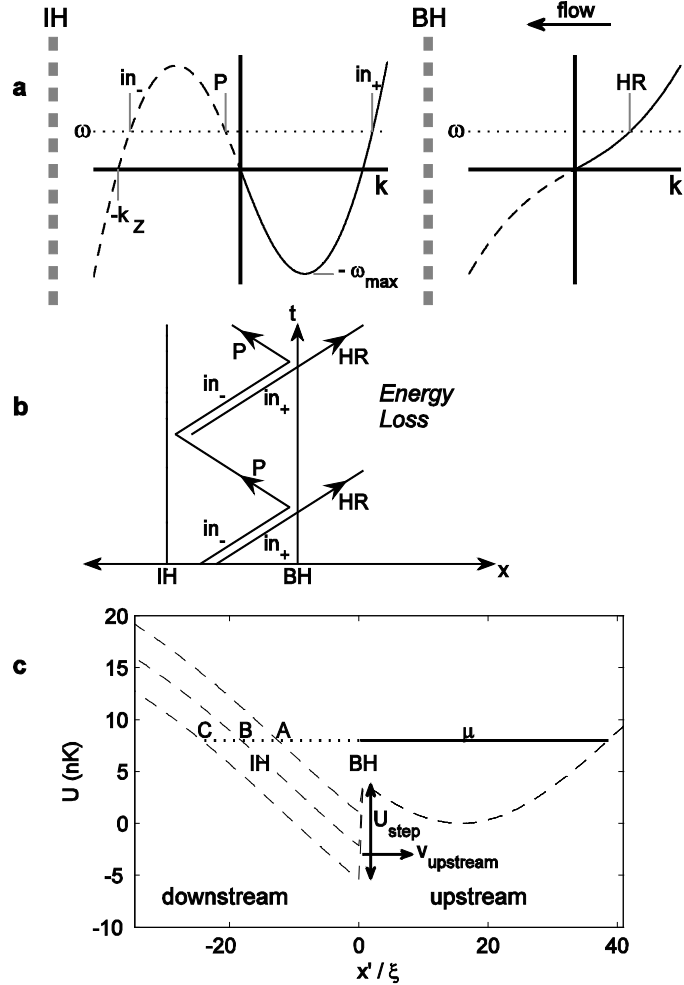

**Figure 1 | The black hole lasing phenomenon and the experimental technique.** The black hole and inner horizons are indicated by *BH* and *IH*, respectively. The Hawking radiation is indicated by *HR* and the negative energy partners of the Hawking radiation are indicated by *P*. **a**, The dispersion relations. The region to the right of *BH* is subsonic. The region between *BH* and *IH* is supersonic. **b**, Spacetime diagram. Energy is lost due to Hawking radiation. **c**, The experiment as viewed in the laboratory frame. The 3 potential profiles employed are indicated by dashed curves. The condensate is accelerated as it flows over the potential step, resulting in the horizon *BH*. The potential minimum to the right of *BH* is stationary, and the potential step moves with speed $v_{\text{upstream}}$. The horizontal line indicates the chemical potential $\mu$ of the condensate. *A*, *B*, and *C* indicate the turning points for the 3 step heights $U_{\text{step}}$. As the turning

point is approached, the flow velocity drops below the speed of sound, resulting in the horizon *IH*. The spatial coordinate $x'$ is given in units of the healing length $\xi$.

Let us consider an initial time with some population of the two modes *in₋* and *in₊* in Fig. 1b. When these modes strike the black hole horizon, they will produce partner particles *P* with negative energy, as well as positive energy Hawking radiation *HR*. This is the Hawking pair production process. Note that the Hawking radiation *HR* has carried away positive energy from the lasing region. Thus, the amplitude of the negative energy modes *P* and *in₋* must increase. The *P* mode then travels to the inner horizon, where it is reflected in the form of *in₋* and *in₊* modes. The process repeats as these modes travel to the black hole horizon, emitting more Hawking radiation and thus further increasing the amplitude of the negative norm modes.

Thus, the standing wave pattern seen in Fig. 2 is a negative energy pattern, resulting from the interference between the Hawking partner mode *P* and *in₋* [36]. The pattern grows as the energy becomes more negative, since positive energy is lost in the form of Hawking radiation. The exponential growth time constant $\tau_L$ of this lasing is proportional to the time $\tau_{RT}$ between emissions of Hawking radiation *HR*. The latter is the round-trip time for excitations to travel from the black hole horizon to the inner horizon and back. We find that these two time constants are related by

$$\tau_L = \frac{\tau_{RT}}{\ln(|\beta(\omega)|^2+1)} \tag{1}$$

where $\beta(\omega)$ is the mixing coefficient giving the efficiency with which Hawking radiation of frequency $\omega$ is produced at the black hole horizon. We have neglected the mixing at the inner horizon since the potential gradient is much smaller there, as illustrated in Fig. 1c. In this work we measure both of the time constants and thus obtain $|\beta(\omega)|^2$, the average number of Hawking particles produced for each *in₋* particle which strikes the horizon [9].

The black hole horizon only strongly mixes modes for which $\hbar\omega$ is less than the Hawking temperature

$$k_B T_H = \frac{\hbar}{2\pi}\left(\frac{dc}{dx} + \frac{dv}{dx}\right), \tag{2}$$

where the derivatives of the flow velocity $v$ and the speed of sound $c$ give the curvature of the horizon, in analogy with the surface gravity. This frequency cutoff is seen in the expression for the mixing coefficient[45],

$$|\beta(\omega)|^2 = \frac{1}{e^{\hbar\omega/k_B T_H} - 1}. \tag{3}$$

This expression is valid as long as $\hbar\omega_{\max}$ seen in Fig. 1a is much larger than $k_B T_H$ [45]. This criterion is well-realized in this experiment since the flow velocity is significantly larger than the speed of sound. By measuring $|\beta(\omega)|^2$, we therefore obtain the ratio between the frequency of the lasing mode and the Hawking temperature.

The mixing coefficient in (3) is seen to diverge for small frequency. Thus, the lasing time constant $\tau_L$ goes to zero for small frequency, by (1). Therefore, the mode with the lowest frequency will dominate the lasing. Furthermore, lower frequency $\omega$ indicated by the dotted line in Fig. 1a corresponds to larger lasing wavenumber $k_P - k_{in-}$ [36]. This implies that the lasing wavenumber is very close to $k_Z$ in Fig. 1a. The latter is given by[46]

$$k_Z \xi = 2\sqrt{\frac{v_L^2}{c_L^2} - 1} \tag{4}$$

where $\xi = \hbar/mc_L$ is the healing length, $v_L$ and $c_L$ are the flow velocity and speed of sound in the lasing region, and $m$ is the atomic mass. We can also compute the group velocities in Fig. 1a for $\omega \approx 0$. This gives us the speed of the partner particle as it propagates the distance $L$ from the black hole horizon to the inner horizon, as well as the speed back to the black hole horizon by the *in-* and *in+* modes. We thus find a round-trip time of

$$\tau_{RT} = L \frac{c_L - 2v_L}{v_L^2 - c_L^2}. \tag{5}$$

The discussion above is valid for any $\omega$. However, the finite length $L$ of the lasing region puts a constraint on the lasing wavenumber $k_P - k_{in-}$ [36]. This results in a discrete spectrum for $\omega$. The calculations of Ref. 36 show that the lowest $\omega$ has the shortest $\tau_L$ and thus dominates the lasing, which agrees with our analysis above.

The condensate forming the black hole laser is confined in space by a focused laser beam (5 μm waist, 123 Hz radial trap frequency, 812 nm wavelength). The confining potential is so weak that it can only provide 9% of the force of gravity. Thus, gravity is compensated by a vertical magnetic field gradient. Initially, the condensate of $^{87}$Rb atoms in the $F = 2$, $m_F = 2$ state is created in a magnetic trap. The condensate is then transferred to the confining beam with sufficient power to support the atoms against gravity. The magnetic field gradient is then ramped on. The power of the confining beam is then ramped down to its final value during 1 s. The beam constricts the condensate to a long tube-like volume. The transverse confinement has an energy level spacing of 6 nK. The chemical potential of the condensate is $\mu = 8$ nK. Thus, the dynamics are nearly one-dimensional, which enhances the visibility of the Hawking radiation[47]. The axial confinement is weaker than parabolic due to the nature of the Gaussian laser beam, as seen in Fig. 1c.

The black hole horizon is created by an additional step-like "waterfall" potential which accelerates the condensate to supersonic speeds, as shown in Fig. 1c. This potential is created by a large-diameter Gaussian laser beam, half of which is blocked by a knife edge. This knife edge is imaged onto the condensate using the same high-resolution optics used for acquiring the images. The illuminated region attracts the atoms, so the sharp transition from dark to light corresponds to a potential step. Since the optical resolution is 1 µm, this step has a width on the order of the healing length ($\xi = 2$ µm in the lasing region). The step potential is swept along the axial length of the condensate by means of a high-optical-resolution acousto-optic modulator (Isomet OPP-834). Thus, the horizon moves at the constant subsonic speed $v_{\text{upstream}} = 0.21$ mm sec$^{-1}$ in the laboratory frame, while the confining potential of the condensate remains at rest. To the right of the step potential, the condensate is essentially unperturbed and at rest in the laboratory frame. To the left of the potential drop however, the condensate flows at supersonic speed. The location of the step is therefore the black hole horizon, indicated in the experimental image of Fig. 2e. It is useful to consider the rest frame of this horizon. In this frame, the region to the right of the step is the upstream region, which flows to the left at the constant applied speed $v_{\text{upstream}}$. The three step heights $U_{\text{step}}$ employed in this work are shown in Fig. 1c. Due to the profile of the confining potential, the flow speed to the left of the black hole horizon decreases as the atoms flow "uphill" and approach the relevant turning point *A*, *B*, or *C*. The point where the flow drops below the speed of sound is the inner horizon, which occurs before the turning point is reached.

The black hole laser is imaged with phase contrast imaging. A very small detuning of one linewidth is employed to maximize the signal. Indeed, the imaging is more sensitive than absorption imaging and has less shot noise. An ensemble of approximately 80 images is collected for a given time. Fig. 2a-g show the average of the images at each time. The lasing mode is visible, particularly at the latest time seen in Fig 2g-i. From Fig. 1c, the width *L* of the supersonic region between the horizons should increase with $U_{\text{step}}$, since the distance between *BH* and *A*, *B*, or *C* increases. This variation in the width is apparent in Fig. 2g-i. Furthermore, increasing $U_{step}$ should increase $v_L$. This increases the lasing wavenumber by (4), as seen in Fig. 2g-i.

By comparing the various times of Fig. 2a-g and applying the continuity equation, we can extract $v$ [41]. From the density, we can compute $c$. The result is shown in Fig. 3a. The supersonic lasing region ($-v > c$) is clear in the figure.

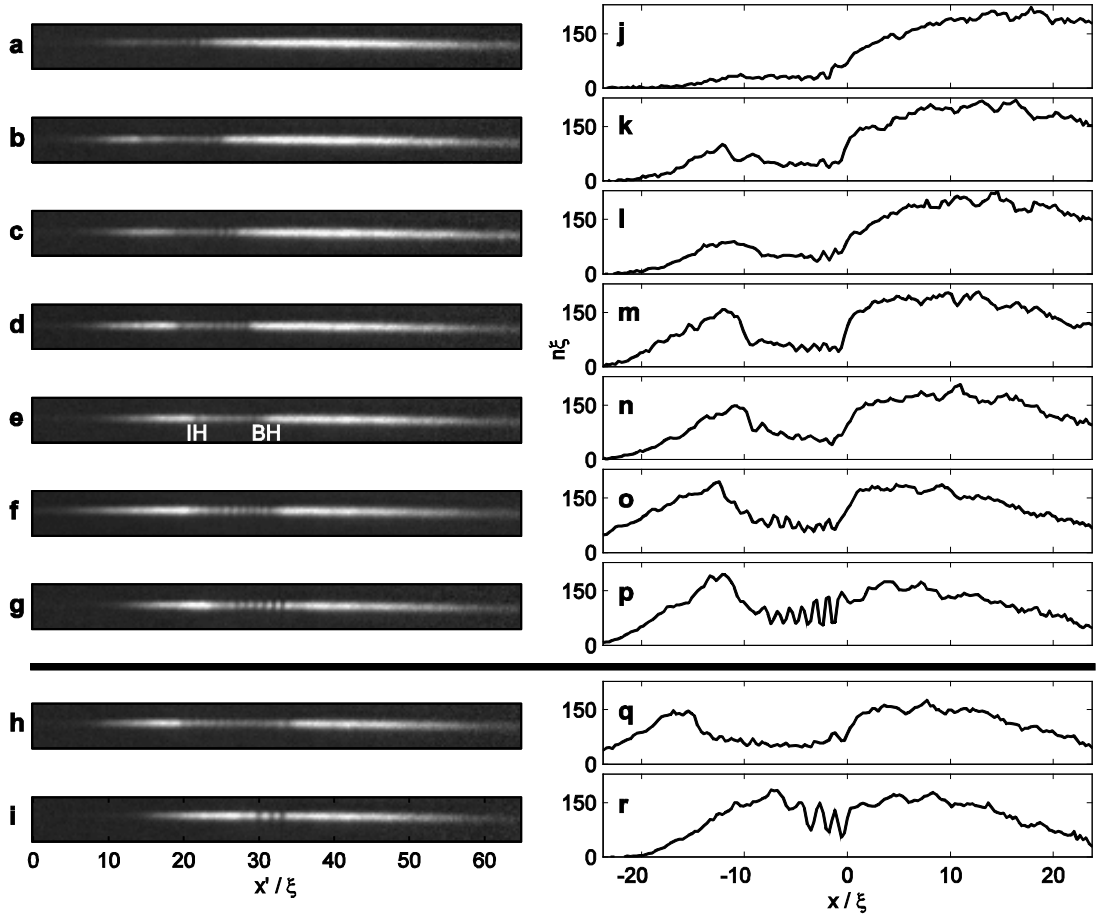

**Figure 2 | The black hole laser. a-g**, *In situ* images at 20 ms time intervals, starting soon after the creation of the black hole and inner horizons. The images show the average of the ensemble as viewed in the laboratory frame, in which the horizons move. The intermediate $U_{step}$ is employed. The spatial coordinate is given in units of the healing length $\xi$. **h**, Like the latest time g, but with the larger $U_{step}$. **i**, Like g, but with the smaller $U_{step}$. **j-r**, Integrated profiles corresponding to a-i, as viewed in the horizon frame. The black hole horizon is located at the origin.

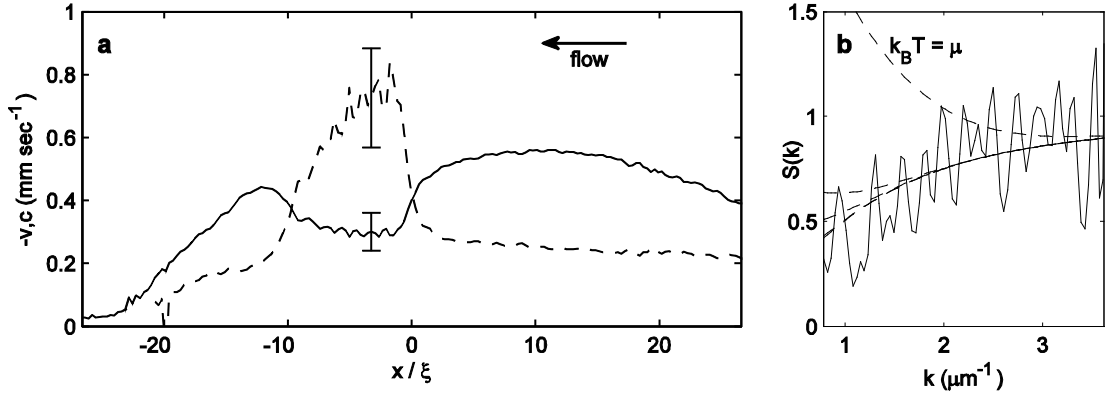

**Figure 3 | The flow velocity, the speed of sound, and the initial temperature. a**, The dashed and solid curves indicate $-v$ and $c$, respectively. The curves indicate the average over the 120 ms corresponding to Figs. 2a-g. **b**, The static structure factor of the upstream region. The solid curve is the measurement with no free parameters. Values less than unity correspond to quantum fluctuations. The theoretical dashed curves correspond to various temperatures. From highest to lowest, the dashed curves correspond to $k_B T = \mu, 0.3\mu, 0.2\mu, 0.1\mu$, and 0.

We would like to determine the initial phonon state of the condensate. We therefore study the condensate upstream for the earliest time studied, which is similar to the condensate before the creation of the horizons. Since the condensate is so cold due to its shallow confining potential, the temperature cannot be determined by the usual time-of-flight technique. Indeed, no thermal atoms are detectable in time-of-flight. We thus employ the technique which we recently developed, in which we observe the fluctuations *in situ*[43]. We compute the Fourier transform $\rho_k$ of the upstream region of each 1D density profile $n(x)$. We compute the static structure factor of the ensemble given by $S(k) = N^{-1}(\langle |\rho_k|^2\rangle - |\langle \rho_k\rangle|^2)$, where $N$ is the number of atoms in the upstream region. The static structure factor is the power spectrum of the fluctuations. The result is shown in Fig. 3b. The result is strikingly quantum, in that $S(k)$ decreases for decreasing $k$. This is in sharp contrast to a state dominated by thermal fluctuations, in which $S(k)$ is proportional to the Planck distribution $[exp(\hbar\omega_k/k_B T) - 1]^{-1}$. The suppression of the fluctuations for small $k$ is due to quantum pair correlations in the ground state. This suppression is seen for the entire accessible $k$-range shown in Fig. 3b. By comparing with the theoretical curves in the figure, we see that the temperature is somewhere between zero and approximately $0.1\mu$. To put this in perspective, $0.1\mu$ corresponds to a thermal fraction of uncondensed atoms of $2 \times 10^{-5}$.

In Fig. 2, the averaging over the ensemble greatly decreases the visibility of the lasing mode. We can see the full visibility and obtain quantitative information by studying the density-density correlation function, given by[46]

$$G^{(2)}(x_1, x_2) = n_L^{-2}[\langle n(x_1)n(x_2)\rangle - \langle n(x_1)\rangle\langle n(x_2)\rangle - \langle n(x_1)\rangle\delta(x_1 - x_2)]$$

where the average is over the ensemble of images, and $n_L$ is the average density in the lasing region. The result is shown in Fig. 4. This correlation pattern is very similar to the prediction of Ref. 36. The lasing region between *BH* and *IH* is characterized by a checkerboard pattern. This pattern implies that the fluctuations have a well-defined wavelength, as well as well-defined nodes. However, close to the inner horizon in the lasing region, the pattern appears to be more similar to fringes parallel to the diagonal rather than a checkerboard. Such fringes imply that the positions of the nodes vary. This occurs because the position of the inner horizon is determined by hydrodynamics, and thus varies slightly from realization to realization. The location of the black hole horizon, in contrast, is determined by the applied potential step, which is the same in every realization. The number of maxima in the lasing mode is 3, 7, and 10 for increasing values of $U_{\text{step}}$, as seen in Figs. 4h-j. Since the dominant lasing mode is the mode with the largest wavenumber $k_P - k_{in-}$, the number of maxima is also the number of modes in the laser cavity.

The lasing mode is seen to extend in vertical and horizontal bands beyond the horizons, into the subsonic regions. The band outside of the black hole is of particular interest, as it represents the escaping flux, the output of the black hole laser[36]. This is the Hawking radiation emitted by the black hole, as indicated by *HR* in Fig. 4g.

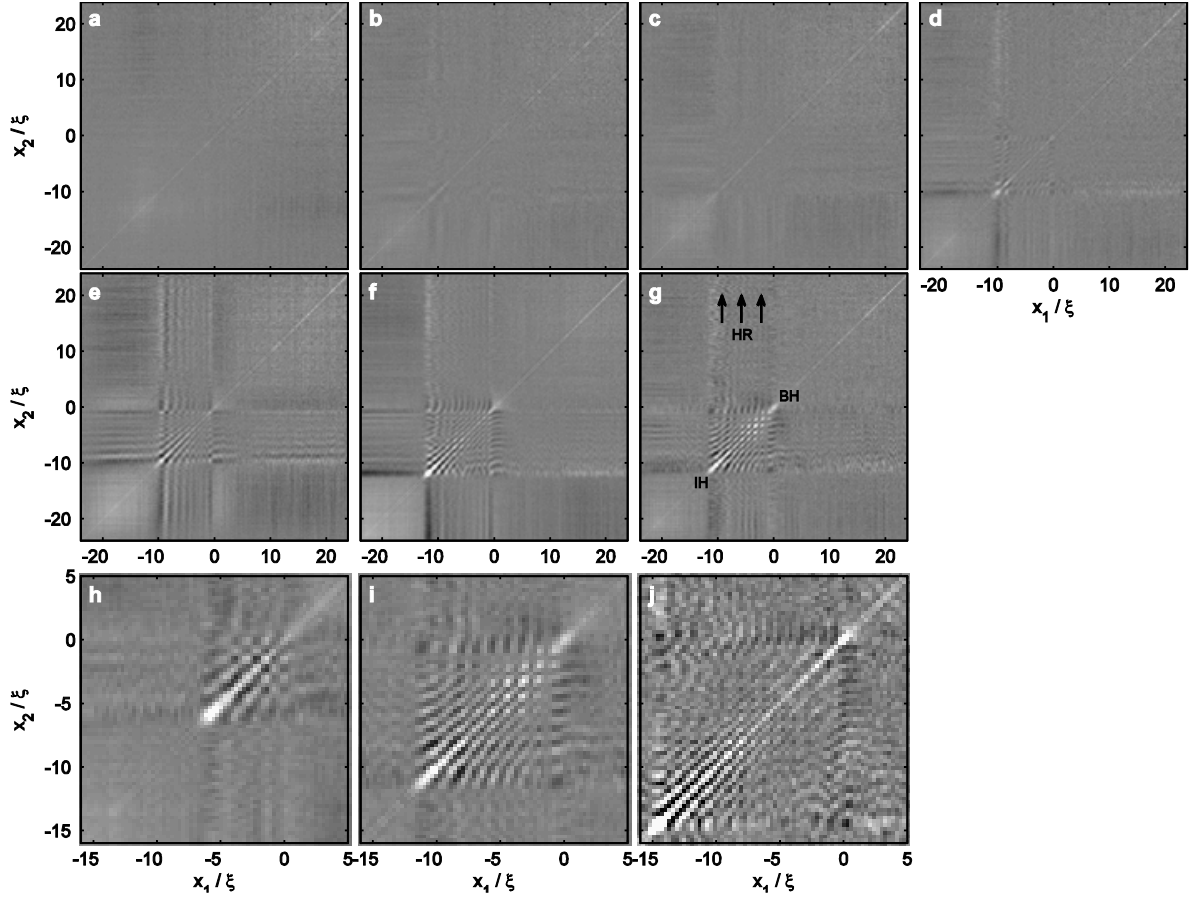

**Figure 4 | Self-amplifying Hawking radiation.** The density-density correlation pattern is shown. The points where $x_1$ and $x_2$ correspond to horizons *BH* and *IH* are indicated in g. The correlations showing the outgoing flux of Hawking radiation are indicated by *HR*. **a-g,** Increasing times corresponding to Fig. 2a-g. The intermediate value of $U_{\text{step}}$ is shown. **h-j,** The lasing region for increasing values of $U_{\text{step}}$ at the latest time. Panel i is an enlargement of g. In j, the grayscale has been decreased relative to h and i by a factor of 2 to improve the visibility of the pattern.

In order to determine the amplitude of the lasing mode, we compute the Fourier transform of the correlation function. This analysis is performed in the region of the correlation function which has a clear checkerboard pattern. In this region the Fourier transform has a clear peak which can be identified and measured. The result is shown in Fig. 5 as a function of time. Each point in the figure corresponds to one panel of Fig. 4. The scale of Fig. 5 is logarithmic, so the linear graph implies exponential growth of the lasing mode. The exponential growth is one way of distinguishing black hole lasing from the phenomena of Hawking radiation from one black hole horizon (no growth), from white hole radiation (linear or logarithmic growth)[46], or from white hole undulations (no growth)[46,48]. The slope of the linear fit is the measured $\tau_L^{-1}$. This time constant is shown in Fig. 6a. The measured values are of the order of magnitude predicted in Ref. 36.

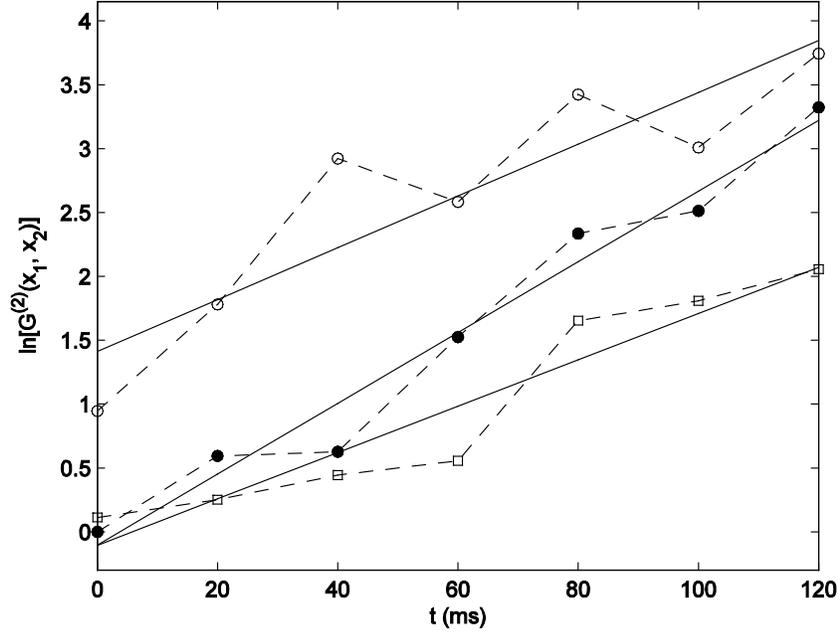

**Figure 5 | The exponential growth of the lasing mode.** The vertical scale is logarithmic. The open circles, filled circles, and squares indicate small, medium, and large $U_{\text{step}}$, respectively. The solid lines are linear fits, corresponding to exponential growth.

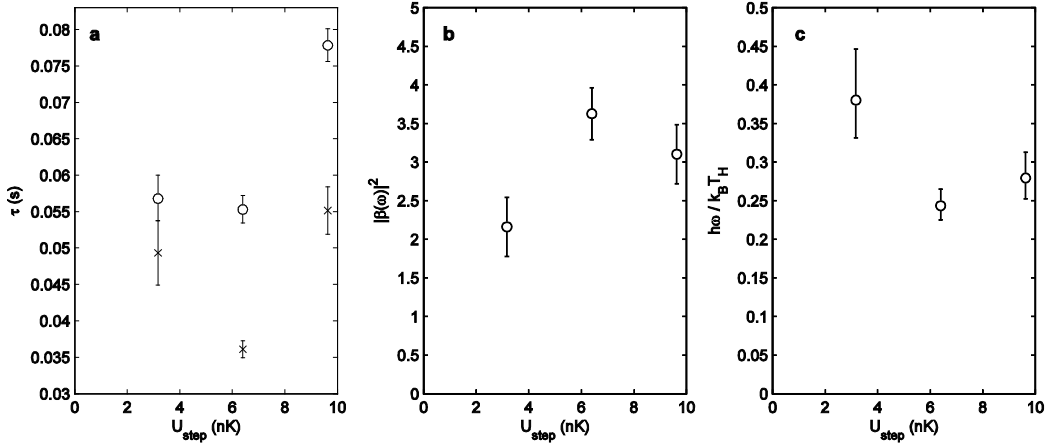

**Figure 6 | The time constants, Hawking particle production coefficient, and lasing energy. a**, The time constants. The lasing time constant $\tau_L$ is indicated by x's. The round-trip time $\tau_{RT}$ is indicated by open circles. They are shown as a function of the height of the potential step of Fig. 1c. **b**, The average number of Hawking particles produced per negative energy particle which strikes the black hole horizon. **c**, The ratio of the lasing frequency to the Hawking temperature.

We can also extract the round-trip time of the excitations. The wavenumber $k_P - k_{in-}$ of the lasing mode is obtained from Figs. 2p-r. The speed of sound $c_L$, and thus the healing length $\xi$, are taken from Fig. 3a. The flow velocity $v_L$ is thus obtained by (4). The length $L$ of the lasing region is clear from Figs. 4h-j. The round-trip time $\tau_{RT}$ is thus found by (5), and is shown in Fig. 6a. The ratio of the two times gives $|\beta(\omega)|^2$ by (1), as shown in Fig. 6b. It is seen that approximately 3 Hawking particles are produced for each incident negative energy particle. This can be translated into the ratio of the frequency of the lasing mode to the Hawking temperature $\hbar\omega/k_B T_H$ by (3), as shown in Fig. 6c.

In conclusion, we have observed self-amplifying Hawking radiation. The emitted Hawking radiation is clearly visible in the density-density correlation function. The lasing mode between the horizons is clear in both the average image as well as the correlation function. The negative energy particles are seen to lase with exponentially increasing amplitude. The frequency of the lasing mode is found to be about 0.3 of the Hawking temperature. This is reasonable in the sense that the horizon should only mix modes with frequency less than the Hawking temperature. This work suggests a method for probing the inside of a black hole. Specifically, if strong laser-like Hawking radiation emanating from an astrophysical black hole were observed, it would imply the existence of an inner horizon. Even more interestingly, it would imply a superluminal dispersion relation. The analog black hole laser is achieved in a weakly trapped, low density, elongated Bose-Einstein condensate with very low temperature, which is floating on a magnetic field gradient. The sharp black hole horizon is created by high-resolution optics. This configuration could be used for additional analog gravity experiments, such as simulating the expansion of the early universe[49-51].

We thank Renaud Parentani, Iacopo Carusotto, Amos Ori, and Florent Michel for helpful discussions. This work is supported by the Russell Berrie Nanotechnology Institute and the Israel Science Foundation.